\documentclass{amsproc}

\newtheorem{theorem}{Theorem}[section]

\theoremstyle{definition}

\theoremstyle{remark}

\numberwithin{equation}{section}

\begin{document}

\title[Resonances on quantum graphs] {Equivalence of resolvent
and scattering resonances on quantum graphs}

\author{Pavel Exner}
\address{Doppler Institute for Mathematical Physics and Applied
Mathematics, B\v{r}ehov\'{a} 7, 11519 Prague, or 25068 \v{R}e\v{z}
near Prague, Czechia} \email{exner@ujf.cas.cz}
\thanks{The research was partially supported by
ASCR and MEYS of the Czech Republic within the projects IRP
AV0Z10480505, A100480501, and LC06-002.}

\author{Ji\v{r}\'{\i} Lipovsk\'{y}}
\address{Institute of Theoretical Physics, Faculty of Mathematics and
Physics, Charles University, V Hole\v{s}ovi\v{c}k\'ach 2, 18000
Prague, Czechia} \email{JLipovsky@seznam.cz}

\subjclass{Primary 81U99, 81V99; Secondary 34L25, 34G10}

\dedicatory{This paper is dedicated to Jean-Michel Combes on the
occasion of his 65th birthday.}

\keywords{Differential geometry, algebraic geometry}

\begin{abstract}
We discuss resonances for Schr\"odinger operators on metric graphs
which consists of a finite compact part and a finite number of
halflines attached to it; the vertex coupling is assumed to be of
the $\delta$-type or certain modifications of it. Using exterior
complex scaling on the graph we show that the resolvent and
scattering resonances coincide in this case.
\end{abstract}

\maketitle

\section{Introduction}

Resonances belong to the class of phenomena which are easy to be
understood on a heuristic level but more difficult when we try to
study them rigorously. It is not a lack of a precise definition,
of course, rather the fact that there are several formal ways in
which the problem can be approached. The two oldest and most
common concepts are \emph{scattering resonances} and
\emph{resolvent resonances}. In the first case one inspects the
on-shell scattering matrix looking for its sharp changes in some
parts of the energy axis suggesting a locally enhanced time delay,
while in the second one we inspect analytical continuation of the
resolvent to the ``unphysical'' sheet(s) and look for poles there.
The two types of resonances need not be \emph {a priori}
identical, and if fact, there is no reason why they should be,
because the latter represent a property of the Hamiltonian $H$ of
the system alone, while the scattering refers to a pair $(H,H_0)$
depending thus on the choice of the free operator. Nevertheless,
in most systems the scattering resonances referring to a
``natural'' $H_0$ coincide with the resolvent ones, which is a
property to be verified in each particular case.

In this paper we are going to address the question of coincidence
between scattering and resolvent resonances in noncompact quantum
graphs. These systems attracted a lot of attention recently from
several reasons, the chief ones among them being that they can be
used to model a wide family of semiconductor and other
microstructures, and that at the same time they represent a
testing ground for investigation of fundamental effects such a
quantum chaotic behavior; we refer to \cite{Ku} for a review and
an extensive bibliography to these problems.

From the mathematical point of view quantum graphs represent
systems of ODE's coupled by boundary conditions, so the question
posed above can be answered in a straightforward way, because the
resolvent of a graph Hamiltonian can be constructed by means of
Krein's formula. In reality, however, such a head-on approach
would be quite laborious because finding the resolvent is not easy
unless the graphs in question has a trivial topology. Fortunately,
there is a simple way: using a powerful insight of Jean-Michel
Combes and collaborators, first formulated in the paper \cite{AC},
one can transform the search for poles of the analytically
continued resolvent into the spectral problem for a suitable
non-selfadjoint operator obtained by an exterior complex scaling
of the Hamiltonian. For graphs with $\delta$-coupling at the
vertices the goal can be be achieved by combing this method with
the well-known duality property \cite{E1}. We will first remind
some needed notions and illustrate our program on examples, then
in Section~3 we will formulate and prove our main result.

\section{Preliminaries}

\subsection{Quantum graphs} \label{qgraphs}

The configuration space of our model is a graph with a finite
number set $\mathcal{V} = \{\mathcal{X}_j:\: j\in I\}$ of
vertices, where $I$ is a corresponding index set, and a set of
edges attached to them; the graph is supposed to be \emph{metric}
so the edges are identified with segments of the real axis. In
particular, vertices connected to $\mathcal{X}_j$ by an edge form
the set of \emph{neighbors} $\mathcal{N}(\mathcal{X}_j) =
\{\mathcal{X}_n:\: n\in \nu(j)\subset I\backslash\{j\}\}$. If
necessary we may assume that any two vertices are connected with
at most one (finite-length) edge, since otherwise we can add a
vertex to each ``superfluous'' edge. With this convention, each
finite-length is uniquely characterized by a pair of vertices, so
we can write their set as ${\mathcal L} =\{\mathcal{L}_{jn}:
(\mathcal{X}_j, \mathcal{X}_n)\in I_\mathcal{L} \subset I\times
I\}$. We will assume that a semi-infinite edge is attached to some
vertices, denoting the graph and its part with edges of finite
lengths as $\Gamma_e$ and $\Gamma$, respectively. The graph
\emph{boundary} $\mathcal{B}$ is the set of vertices which have a
single neighbor, the \emph{ interior} is $\mathcal{I} =
\mathcal{V}\backslash\mathcal{B}$. We denote by $\mathcal{C}$ the
set of vertices to which a semi-infinite link is attached, and we
introduce $I_\mathcal{B}$, $I_\mathcal{I}$,  $I_\mathcal{C}$,
respectively, as the appropriate index subsets in $I$.

Using the metric structure of our graph we can introduce the state
Hilbert space of the problem as the $L^2$ space on $\Gamma_e$
which is naturally identified with
 $$
    \mathcal{H} = \bigoplus_{(j,n) \in I_\mathcal{L}}
    L^2([0,l_{jn}])\bigoplus_{j\in I_\mathcal{C}}
    L^2([0,\infty))\,,
 $$
its elements being written by $\psi  = \{\psi_{jn}: (j,n)\in
I_\mathcal{L},\, \psi_{j\infty}: j\in I_\mathcal{C}\}$. The
Hamiltonian of the model is a Schr\"odinger operator on $\Gamma_e$
which acts as $-\frac{\mathrm{d}^2}{\mathrm{d}x^2} +V_{jn}(x)$ at
each edge with a family of potentials $\{V_{jn},V_{j\infty}\}$ and
appropriate boundary conditions at the vertices. To make things
simple we will suppose that $V_{jn} \in L^{\infty}([0, l_{jn}])$,
and moreover, that the motion is free at each external link,
$V_{j\infty} = 0$. The describe the boundary conditions we need
boundary values. Identifying $\mathcal{X}_j$ with $x = 0$ we
introduce $\psi_{jn} = \lim_{x\to 0+} \psi_{jn}(x)$ and
$\psi_{jn}' = \lim_{x\to 0+} \psi_{jn}'(x)$, and arrange them as
into columns, $ \Psi_j= (\psi_{jn}(\mathcal{X}_j):\: n\in
\nu(j))^T$, and similarly for $\Psi'_j$. The boundary condition at
the vertices must ensure that the Schr\"odinger operator on
$\Gamma_e$ is self-adjoint; it is well known \cite{KS1} that this
happens if
 \begin{equation} \label{bc}
    A_j \Psi_j + B_j \Psi'_j = 0\,, \quad j\in I\,,
 \end{equation}
where $A_j,\,B_j$ are matrices of $d_j:= \mathrm{card}\,\nu(j)$
such that $(A_j,B_j)$ has maximal rank  $A_j B_j^*$ is
self-adjoint at each vertex. Moreover, the non-uniqueness in the
above conditions can be removed if one chooses $A_j+U_j-I,\, B_j=
i(U_j+I)$ where $U_j$ is a $d\times d$ unitary matrix \cite{Ha,
KS2}. Hence the family of admissible is characterized by $d_j^2$
real parameters for each vertex. In this large set there are some
distinguished case. One of them is the \emph{$\delta$-coupling} for
which the wave functions are continuous at the vertex,
$\psi_{jn}(\mathcal{X}_j) = \psi_{jm}(\mathcal{X}_j)=: \psi_j$,
and $\sum_{n\in\nu(j)} \psi_{jn}' (\mathcal{X}_j) = \alpha_j
\psi_j$. Another interesting case is the
\emph{$\delta'_s$-coupling} defined by similar conditions with the
roles of functions and derivatives interchanged; other examples
will be mentioned below.

\subsection{Exterior complex scaling}

A change of coordinates gives means generally to replace the
Hamiltonian by a unitarily equivalent operator. The basic idea of
the method mentioned above is to use a family of such
transformation, corresponding to scalings in the whole space or in
the exterior of a given domain, and to extend it analytically to
complex values of the scaling parameter. This typically leads to a
non-selfadjoint operator with the essential spectrum rotated; by a
suitable parameter choice the resonance poles can be then found as
complex eigenvalues of the transformed operator, corresponding to
eigenfunctions in the Hilbert space.

For our system a natural idea is to employ a family of
transformations which leaves the compact part $\Gamma$ of
$\Gamma_e$ intact and \emph{scales the external semi-infinite
edges.} Identifying such a link with the halfline $\mathbb{R}^+$
we consider the scaling transformation $g_\theta \to U_\theta g=
\mathrm{e}^{\theta/2}g(x\mathrm{e}^{\theta})$ with the
parameter $\theta$. Application of $U_\theta$ to the Laplacian
leads to its multiplication by $\mathrm{e}^{-2\theta}$, on the
halfline one has naturally to transform also the boundary
condition. Scaling the external edges, all with same parameter, we
replace therefore the original graph Hamiltonian by the operator
 $$
    H_\theta {\{g_j\} \choose \{f_{jn}\}}
    = {\{-\mathrm{e}^{-2\theta} g_j''\} \choose
    \{-f_{jn}'' + V_{jn} f_{jn}\}}
 $$
where the upper component corresponds to the external edges, $g_j$
being the wave function on the halfline attached to the vertex
$\mathcal{X}_{j}$, and $\{f_{jn}\}$ corresponds similarly to
interior edges of the graph. The domain of the transformed
operator consists of functions with components $f_{jn} \in
W^{2,2}([0,l_{jn}])$ and $g_{j\theta} = U_\theta g_j $ with
$g_j\in W^{2,2}(\mathcal{L}_{j\infty})$ satisfying the
appropriately transformed boundary conditions. We will be
interested in the nontrivial situation when $\theta$ is complex,
for instance $\theta =i\vartheta$ with $\vartheta>0$. The essential
spectrum of $H_\theta$ comes clearly from the external edges and
the above formula shows that it is rotates into the lower complex
halfplane; for $\vartheta$ large enough one can uncover the poles of
$H$ laying on the second sheet.

Before we pass to formulating the general result, let us discuss
several examples.

\subsection{Example: a line with an appendix} \label{e: append}

The simplest nontrivial example consist a line, representing to
external links, to which a line segment of length $l>0$ is
attached at the point $x=0$. Consequently, the Hilbert space is
$L^2(\mathbb{R})\oplus L^2([0,l])$, its elements can be written as
$\psi = {g\choose f}$ where $g,f$ refers to the line and the
appendix, respectively, and the Hamiltonian acts as $H \psi =
{-g''\choose -f''+Vf}$. To make it self-adjoint we have to choose
boundary conditions at the point $x=0$ and the other endpoint of
the segment. Let the latter be Dirichlet, $f(l)=0$, and at the
vertex we put
 $$
   g(0+) = g(0-)=: g(0), \quad f(0) = \beta g(0)+ \gamma f'(0),
   \quad g'(0+) - g'(0-)= \delta g(0)-\beta f'(0)
 $$
for some $\beta, \gamma, \delta \in\mathbb{R}$ following
\cite{ES}; it is the most general class of time-reversal
Hamiltonians with the line wave function continuous at $x=0$, in
particular, the case $\beta=1$, $\gamma=0$ represents the
$\delta$-coupling with $\alpha=\delta$. The scattering problem for this
system is easy to solve \cite{ES}: denoting by $f_l(x)$ the
normalized Dirichlet solution to the Schr\"odinger equation at the
appendix, we can express the reflection and transmission
amplitudes at energy $k^2$ as
 $$
  r = \frac{\delta[f_l(0)-\gamma f_l'(0)]-\beta^2f_l'(0)}
  {(2ik-\delta)[f_l(0)-\gamma f_l'(0)]+\beta^2 f_l'(0)}\,,\quad
  t = \frac{2ik[f_l(0)-\gamma f_l'(0)]}{(2ik-\delta)
  [f_l(0)-\gamma f_l'(0)]+\beta^2f_l'(0)} \,.
 $$
Scattering resonances coincides with zeros of the denominator;
their behavior with respect to the parameters is discussed in
\cite{ES}. If we suppose for simplicity that $V=0$, then
$f_l(x)=\sin{k(l-x)}$ and the condition can be rewritten as
\begin{equation} \label{appendres}
    \tan{kl}=\frac{\beta^2 k}{2ik-\delta}-\gamma k.
\end{equation}
Let us look now at the problem from the complex scaling point of
view and put $g_\theta (x) := \mathrm{e}^{\theta/2}
g(\mathrm{e}^{\theta}x)$. The corresponding boundary values are
$g_\theta(0)= \mathrm{e}^{\theta/2}g(0)$ and $g_\theta'(0\pm)=
\mathrm{e}^{3\theta/2} g'(0\pm)$. This has to be
substituted into the above boundary conditions: the continuity at
$x=0$ and the Dirichlet condition at the appendix endpoint do not
change, while the other two yield
 $$
  f(0)= \beta\mathrm{e}^{-\theta/2}g_\theta(0)+\gamma f'(0)\,,\;
  \mathrm{e}^{-3\theta/2}[g_\theta'(0+)-g_\theta'(0-)] =
  \delta\mathrm{e}^{-\theta/2}(0)g_\theta(0)-\beta f'(0)\,;
 $$
this can be regarded as the boundary conditions which define the
non-selfadjoint operator $H_\theta$. Assuming again $V=0$, it
is easy to solve the corresponding spectral problem. The appendix
solution is  $f(x) = b\, \sin{k(l-x)}$ and its halfline
counterparts are the corresponding exponential functions
$\mathrm{e}^{\mp ikx\mathrm{e}^{\theta}}$.
The first one of the above conditions gives
 $$
  b = \frac{\beta\mathrm{e}^{-\theta/2}}{\sin{kl}
  +\gamma k\cos{kl}}\, g_\theta(0)
 $$
and substituting from here into the second one we get
 $$
  \mathrm{e}^{-3\theta/2}g_\theta(0)2ik\mathrm{e}^{\theta} =
  g_\theta(0)\mathrm{e}^{-\theta/2}\left[\delta-\frac{\beta^2(-k\cos{kl})}
  {\sin{kl}+\gamma k\cos{kl}}\right]\,.
 $$
It is straightforward to conclude from the last relation that the
resolvent resonances for the line with an appendix are determined
by the condition (\ref{appendres}) again.

\subsection{Example: a loop with two leads}

As the next example let us consider a graph consisting of two
internal edges of lengths~$l_1,\,l_2$, connecting the endpoints of
two halflines. Consequently, the Hilbert space is
$L^2(\mathbb{R}^-)\oplus L^2(\mathbb{R}^+)\oplus L^2([0,l_1])
\oplus L^2([0,l_2])$ and states are described by the columns
$\psi=(f,g,u,v)^T$, with the Hamiltonian acting as
$\psi=(-f'',-g'',-u'',-v'')^T$; for simplicity we suppose again
that the particle is free away of the vertices. The boundary
conditions will be now chosen as a $\delta$-coupling at each
vertex, i.e. the continuity together with
 $$
  u'(0)+v'(0)-f'(0)= \alpha f(0)\,, \quad
  -u'(l_1)-v'(l_2)+ g'(0) =\beta g(0)\,.
 $$
The scattering problem is again easy to solve. One uses the above
boundary conditions to match the solutions $\mathrm{e}^{ikx} +r\,
\mathrm{e}^{-ikx}$ and $t\,\mathrm{e}^{ikx}$ for $\mp x>0$ with
the linear combinations of the exponentials at the internal links;
solving the corresponding system of linear equations we get the
on-shell reflection and transmission amplitudes
 $$
 r(k) = \frac{i- \gamma(k)}{i+\gamma(k)}\,, \quad
 t(k) = \frac{2i}{\gamma(k) +i}\,,
 $$
where
 $$
 \gamma(k) :=\frac{\left(\frac{1}{\sin{kl_1}}+\frac{1}
 {\sin{kl_2}}\right)^2}{\frac{1}{\tan{kl_1}}+
 \frac{1}{\tan{kl_2}}+\frac{\beta}{k}-i}-
 \left(\frac{1}{\tan{kl_1}}+\frac{1}{\tan{kl_2}}\right)-\frac{\alpha}{k}\,.
 $$
In particular, the scattering resonances are now determined by the
condition
\begin{equation} \label{loopres}
   \gamma(k) +i=0\,.
\end{equation}
Let us now apply the scaling to both halflines putting
$f_\theta(x):= \mathrm{e}^{\theta/2}
f(x\mathrm{e}^{\theta})$ and $g_\theta(x):=
\mathrm{e}^{\theta/2} g(x\mathrm{e}^{\theta})$. Calculating
the new boundary values we can write down the boundary conditions
determining $H_\theta$, the continuity at the vertices together
with
\begin{eqnarray*}
    -\mathrm{e}^{-3\theta/2}f'_\theta(0-) + u'(0)+v'(0)
    = \alpha\mathrm{e}^{-\theta/2}f_\theta (0-), \\
    \mathrm{e}^{-3\theta/2}g'_\theta(0+) - u'(l_1)-v'(l_2)
    = \beta\mathrm{e}^{-\theta/2}g_\theta (0+).
\end{eqnarray*}
To solve the eigenvalue problem for $H_\theta$ one has to match
the solutions at the external edges which are $\mathrm{e}^{\mp
ikx\mathrm{e}^{\theta}}$ with
 $$
 u(x) = \frac{f_\theta (0-)\sin{k(l_1-x)+g_\theta(0+)\sin{kx}}}
 {\sin{kl_1}}\,\mathrm{e}^{-\theta/2}
 $$
and the analogous solution at the other internal link.  This
yields the conditions
\begin{eqnarray*}
k\left[g_\theta(0+)\left(\dfrac{1}{\sin{kl_1}}+
    \dfrac{1}{\sin{kl_2}}\right)-f_\theta(0-)\left(\dfrac{1}{\tan{kl_1}}+
         \dfrac{1}{\tan{kl_2}}\right)\right] &\!\!=\!\!&
    (\alpha\!-\!ik)f_\theta(0-)\,, \\
    k\left[g_\theta(0+)\left(\dfrac{1}{\tan{kl_1}}+
    \dfrac{1}{\tan{kl_2}}\right)-f_\theta(0-)\left(\dfrac{1}{\sin{kl_1}}+
         \dfrac{1}{\sin{kl_2}}\right)\right ] &\!\!=\!\!&\!\!
    -(\beta\!-\!ik)g_\theta(0+)\,,
\end{eqnarray*}
which are obviously equivalent to the scattering-resonance
equation (\ref{loopres}).

\subsection{Example: a magnetic lasso graph} \label{e: lasso}

To show that these consideration extend beyond the pure
Schr\"odinger case, take now a graph consisting of a loop of
circumference $l$ to which a halfline is attached and suppose that
it is placed into a magnetic field perpendicular to the loop
plane. The Hilbert space is thus $L^2(\mathbb{R}^+)\bigoplus
L^2([0,l])$ with elements $\psi = {g\choose f}$ and the
Hamiltonian acts as
 $$
 H \psi = H {g \choose f} =
 {-g'' \choose -f'' -2i A f'+A^2 f}\,,
 $$
where $A$ is the corresponding vector potential. In fact, the form
of the field is not important, what matters is the flux through
the loop. The boundary conditions we consider are similar to those
of Example~\ref{e: append}: we suppose that the wave function is
continuous on the loop, $f(0)=f(l)$, and furthermore,
 $$
 f(0) = \alpha^{-1}[f'(0)-f'(l)]+ \gamma g'(0)\,, \quad
 g(0) = \bar\gamma [f'(0)-f'(l)]+ \tilde\alpha^{-1} g'(0)
 $$
for some $\alpha, \tilde\alpha\in \mathbb{R}$ and $\gamma\in
\mathbb{C}$; now we do not require time-reversal invariance.
Performing the complex scaling on the halfline in the same way as
above, the boundary conditions become
 \begin{eqnarray*}
 f(0) &\!=\!& \alpha^{-1}[f'(0)-f'(l)]+
 \gamma \mathrm{e}^{-3\theta /2} g_\theta'(0)\,, \\
 \mathrm{e}^{-\theta /2 } g_\theta(0) &\!=\!&
 \bar\gamma [f'(0)-f'(l)]+ \tilde\alpha^{-1}
 \mathrm{e}^{-3\theta/2}g_\theta'(0).
 \end{eqnarray*}
To solve the eigenvalue problem for $H_\theta$ we use the
following Ansatz: $g_\theta(x)
=\mathrm{e}^{ik\mathrm{e}^{\theta}}$ and $f(x) = C
\mathrm{e}^{-iAx} \sin{(kx+\varphi)}$ with $\varphi$ given by
$\tan\varphi = \sin kl( \mathrm{e}^{iAx} - \cos kl)$. A
straightforward computation yields then the resonance condition
 \begin{equation} \label{lassores}
    \sin{kl} - 2\left(\frac{k}{\alpha}+\frac{ik^2 |\gamma|^2}
    {1-ik \tilde \alpha^{-1}}\right) (\cos Al-\cos{kl})=0\,,
 \end{equation}
which is again the same as the following from the scattering on
the lasso \cite{E2}.

\section{General graphs}

\subsection{Choice of the vertex coupling} \label{vertex}

In order to extend the observation made in the examples to the
class of graphs described in Sec.~\ref{qgraphs} we have first to
specify the vertex boundary conditions we will consider, in
particular, the way in which the external links are attached to
$\Gamma$. For simplicity we suppose that \emph{at most one}
halfline sprouts of each point of the boundary of $\Gamma$. Let us
denote the wave function on such a halfline referring to
$\mathcal{X}_j\in \mathcal{B}$ as $g_j$, and those on edges
joining $\mathcal{X}_j$ with interior vertices as $f_{jn}\,,
j=1,\dots,m$. To keep things simple we will consider a coupling
which generalizes directly the one of Example~\ref{e: lasso}: the
functions are \emph{continuous} at the vertex, $f_{j1}(0) =
f_{j2}(0) = \ldots = f_{jm} (0) =: f_j(0)$, and
 \begin{equation} \label{extcoupl}
    f_j(0) = \alpha_j^{-1} \sum_{n=1}^m f_{jn}'(0) + \gamma_j
    g_j'(0)\,, \quad
    g_j(0) = \bar\gamma_j\sum_{n=1}^m f_{jn}'(0) + \tilde\alpha_j^{-1} g_j'(0)
 \end{equation}
for $\alpha_j, \tilde\alpha_j\in \mathbb{R}$ and $\gamma_j\in
\mathbb{C}$; it is straightforward to check that it is of the type
(\ref{bc}). In the interior vertices of $\Gamma$ we suppose a
$\delta$ coupling; in the same notation it is \emph{continuity}
again, $f_{j1}(0) = f_{j2}(0) = \ldots = f_{jm} (0) =: f_j(0)$,
and
 \begin{equation} \label{intcoupl}
    \sum_{n=1}^m f_{jn}'(0)= \alpha_j f_j(0)
 \end{equation}
for a real $\alpha_j$, in general different at different vertices.
Finally, if there is a vertex in the boundary of $\Gamma$ to which
no external edge is attached, we assume Dirichlet boundary
conditions there, $f_{jn}=0$ for $j\in I_\mathcal{B}\setminus
I_\mathcal{C}$.

\subsection{A duality}

As we have indicated in the introduction, our second main tool
will be a duality between Schr\"odinger equation,
 \begin{equation} \label{GSE}
 H\psi= k^2\psi\,,
 \end{equation}
on a graph with a $\delta$-coupling and a certain difference
equation. Let us recall it now in more details. To cover both
$\Gamma$ and $\Gamma_e$ it is useful to formulate the result in a
way which allows to describe generalized eigenfunctions at the
same time, hence we consider the class $D_\mathrm{loc}(H)$ which
is the subset in $\bigvee_{(j,n)\in I_\mathcal{L}}
L^2(0,\ell_{jn})\,$ (the direct sum) consisting of the functions
which satisfy all the requirements imposed at $\psi\in D(H)$
except the global square integrability.

On the edge $\mathcal{L}_{nj}\equiv [0,\ell_{jn}]$, with the right
endpoint identified to the vertex $\mathcal{X}_j$, we denote as
$u_{jn}$ and $v_{jn}$ the normalized Dirichlet solutions to
$-f''+U_{jn}f=k^2f$, i.e. those satisfying the boundary conditions
 $$
 u_{jn}(\ell_{jn})= 1\!-\!(u_{jn})'(\ell_{jn})=0\,, \;\; v_{jn}(0)=
 1\!-\!(v_{jn})'(0)=0\,;
 $$
their Wronskian is $W_{jn}= -v_{jn}(\ell_{jn}) =u_{jn}(0)$. Since
in our case the set $\mathcal{V}$ is finite, the assumptions used
in \cite{E1} are satisfied and we have the following result.

 \begin{theorem} \label{duality}
 Suppose that $\psi\in D_{loc}(H)$ solves (\ref{GSE}) for some
$k\not\in\mathcal{K}$ with $k^2\in\mathbb{R},\: \mathrm{Im\,} k\ge
0$. Then the corresponding boundary values satisfy the equation
   \begin{equation} \label{discrete delta}
\sum_{n\in\nu(j)\cap I_\mathcal{I}} {\psi_n\over W_{jn}}\,-\,
\left(\, \sum_{n\in\nu(j)} {(v_{jn})'(\ell_{jn}) \over W_{jn}}
-\alpha_j\, \right)\psi_j= 0\,,
   \end{equation}
and conversely, any solution $\{\psi_j:\, j\in I_\mathcal{I}\}$ of
the system to (\ref{discrete delta}) determines a solution of
(\ref{GSE}) by
   \begin{eqnarray*}
\psi_{jn}(x)= {\psi_n\over W_{jn}}\,u_{jn}(x) -\,{\psi_j\over
W_{jn}}\,v_{jn}(x) \;\;
& {\rm if} & n\in \nu(j)\cap I_\mathcal{I}\,, \\
\psi_{jn}(x)= -\,{\psi_j\over W_{jn}}\,v_{jn}(x) \;\; & {\rm if} &
n\in \nu(j)\cap I_\mathcal{B}\,.
   \end{eqnarray*}
 \end{theorem}
When applied to $\Gamma_e$ the above result concerns also
generalized eigenfunctions. It is useful, however, to specify it
for the scattering situation. Let us consider the solutions on the
external links $g_j(x) = a_j \mathrm{e}^{-ikx}+ b_j
\mathrm{e}^{ikx}$ for all $j\in I_\mathcal{C}$. The operator $S$
maps the vector of incoming amplitudes $a=\{a_j\}$ into the vector
of outgoing amplitudes $b=\{b_j\}$, i.e. $b = Sa$. Poles of the
scattering matrix are given by the condition $\det S^{-1}=0$;
recall that for a nonreal $k$ the matrix ceases to be unitary.
Substituting the Ansatz into the boundary
conditions~(\ref{extcoupl}) we obtain for $j \in I_\mathcal{C}$
  \begin{eqnarray*}
    \alpha_j \psi_j = \sum_{n=1}^m \psi_{jn}'(0) + ik \alpha_j
    \gamma_j (b_j-a_j)\,,\\
    \tilde\alpha_j(a_j + b_j) = \tilde\alpha_j \bar\gamma_j
    \sum_{n=1}^m \psi_{jn}'(0) + ik (b_j-a_j)\,,
  \end{eqnarray*}
while for $j \not\in I_\mathcal{C}$ we have the standard
$\delta$-coupling
  $$
    \alpha_j \psi_j(j) = \sum_{n\in \nu(j)} \psi_{jn}'(j)\,.
  $$
Using Theorem~\ref{duality} we can now proceed in the way similar
to \cite{E2} to obtain the system of equations for $j \in
I_\mathcal{C}$
  \begin{equation}\label{smatrix1}
\alpha_j \psi_j = \sum_{n\in \nu(j)\cap I_\mathcal{I}} -
\frac{\psi_n}{W_{jn}}\,+\,\sum_{n\in \nu(j)}\frac{v_{jn}'(l_{jn})}
{W_{jn}}\,\psi_j + ik \alpha_j\gamma_j (b_j-a_j)\,,
  \end{equation}
  \begin{equation}\label{smatrix2}
\tilde\alpha_j (b_j + a_j) = \tilde\alpha_j \bar\gamma_j \left(\sum_{n\in
\nu(j)\cap I_\mathcal{I}} -\frac{\psi_n}{W_{jn}}\,+\,\sum_{n\in \nu(j)}
\frac{v_{jn}'(l_{jn})}{W_{jn}}\,\psi_j\right) + ik(b_j - a_j)\,.
  \end{equation}
On the other hand, for $j \not\in I_\mathcal{C}$ we have the
condition (\ref{discrete delta}). We have thus arrived at a system
of $N= I+ I_\mathcal{C}$ equations for $\psi_j$ and $b_j$ which
gives, in particular, the sought S-matrix relating the incoming
and outgoing amplitudes.

\subsection{Duality for a complex-scaled graph}

Let us now perform the exterior complex scaling on $\Gamma_e$
which changes the external-edge wave function $g_j$ to
$g_{j,\theta} (x) := \mathrm{e}^{\theta/2}
g_j(\mathrm{e}^{\theta}x)$; the scaling parameter will be the
same for all $j\in I_\mathcal{C}$. Our aim is to find eigenvalues
of the scaled operator $H_\theta$, and since there are no
potentials on the external links, we know that the solutions there
will be of the form $g_{j,\theta}(x) = \mathrm{e}^{ik\mathrm{e}
^{\theta} x} g_{j,\theta}(0)$. Substituting from here to
(\ref{extcoupl}) we get
 $$
 f_j(0) = \alpha_j^{-1}\sum_{n = 1}^{m} f_{jn}'(0)
 + ik\gamma_j \mathrm{e}^{-\theta/2} g_{j,\theta}(0)\,, \quad
 \mathrm{e}^{-\theta/2}g_{j,\theta}(0)
 = \bar\gamma_j\sum_{n=1}^m f_{jn}'(0) + ik\tilde\alpha_j^{-1}\,,
 $$
and eliminating $g_{j,\theta}(0)$ from here we arrive at
 $$
 f_j(0) = \left(\alpha_j^{-1}+\frac{ik |\gamma_j|^2}
 {1-ik \tilde\alpha_j^{-1}}\right) \sum_{n = 1}^{m} f_{jn}'(0)\,.
 $$
In other words, the scaling to replacement of the coupling at the
vertex by a new efficient one, non-selfadjoint and
energy-dependent, with the parameter
 \begin{equation} \label{beta}
 \beta_j (k) := \alpha_j \frac{1-ik \tilde\alpha_j^{-1}}
 {1+ik (|\gamma_j|^2\alpha_j-\tilde\alpha_j^{-1})}\,,
 \end{equation}
which applies only to the interior edges meeting at the vertex
$\mathcal{X}_j$. Now we can repeat step by step the proof of
Theorem~\ref{duality} given in \cite{E1} to find the system of
difference equations determining the eigenfunctions of
$H_\theta$ through their values at the vertices. Those
referring to the interior ones, $j\in I_\mathcal{I}$, do not
change being again given by (\ref{discrete delta}). On the other
hand, the equations for $j\in I_\mathcal{C}$ become
   \begin{equation} \label{modeq}
\sum_{n\in\nu(j)\cap I_\mathcal{I}} {\psi_n\over W_{jn}}\,-\,
\left(\, \sum_{n\in\nu(j)} {(v_{jn})'(\ell_{jn}) \over W_{jn}}
-\beta_j(k)\, \right)\psi_j= 0\,,
   \end{equation}
Now we are ready to compare both systems. Substituting from
(\ref{smatrix1}) into (\ref{smatrix2}) we get after a
straightforward computation
   \begin{equation*}
b_j-a_j=\frac{\alpha_j\bar\gamma_j\psi_j-2a_j}{1+ik(a_j
|\gamma_j|^2-\tilde \alpha_j^{-1})}\,.
   \end{equation*}
Substituting this into (\ref{smatrix1}) again we obtain the system
of equations

 $$
\sum_{n\in\nu(j) \cap I_\mathcal{I}} \frac{\psi_n}{W_{jn}}
-\left(\sum_{n\in\nu(j)} \frac{(v_{jn})'(l_{jn})}{W_{jn}}-\beta_j
(k)\right)\psi_j =\frac{2ik \alpha_j \gamma_j
a_j}{1+ik(|\gamma_j|^2\alpha_j-\tilde\alpha_j^{-1})}
 $$
and (\ref{discrete delta}). As a final step, it is easy to see
that the determinants of both system yield the same pole
condition.

\subsection{The main result}

Summarizing the above discussion we are able now to state the
claim announced in the introduction about the two resonance sets.

 \begin{theorem}
 Let $H$ be a Schr\"odinger operator on $\Gamma_e$ as described in
 Secs.~\ref{qgraphs} and \ref{vertex}, then the families of its
 resolvent and scattering resonances coincide.
 \end{theorem}

\section{Concluding remarks}

The situation when the graph loops are pierced by magnetic fluxes
the example of which we saw in Sec.~\ref{e: lasso} can treated in
the general case also since it reduces to a simple transformation
of wave functions on the internal edges \cite{E1}. Furthermore,
the duality used in the argument is not restricted to the
$\delta$-coupling; in \cite{E1} its validity is demonstrated for
its $\delta'_s$-counterpart. On the other hand, it is natural to
expect that the equivalence of the two resonance types is valid
for any coupling (\ref{bc}), however, to prove this claim the
present approach needs to be modified.

\bibliographystyle{amsalpha}

\end{document}